\documentclass[a4paper]{jpconf}
\usepackage{graphicx}
\usepackage{amsmath}

\begin{document}
\title{Boosted top: experimental tools overview}

\author{Emanuele Usai}

\address{University of Hamburg, on behalf of the ATLAS and CMS collaborations}

\ead{emanuele.usai@desy.de}

\begin{abstract}
An overview of tools and methods for the reconstruction of high-boost top quark decays at the LHC is given in this report. The focus is on hadronic decays, in particular an overview of the current status of top quark taggers in physics analyses is presented. The most widely used jet substructure techniques, normally used in combination with top quark taggers, are reviewed. Special techniques to treat pileup in large cone jets are described, along with a comparison of the performance of several boosted top quark reconstruction techniques.
\end{abstract}


\section{Strategy}
The strategy to address the reconstruction of boosted top quarks can be subdivided in two different categories depending on the decay of the top quark.
For high-$p_T$ top quarks decaying leptonically, the main issue is the small angular separation between the the lepton and the $b$-jet.  Classic lepton isolation algorithm might then reject boosted leptonic top quarks.  A solution to this problem consists in shrinking the isolation cone around the lepton in function of its $p_T$ \cite{atlminiiso}.
For boosted hadronic top quark decays, the main topic of this document, the hadronic decay products might be too close together for standard jet algorithms and sizes to be reconstructed separately. The strategy for the reconstruction of boosted hadronic top quarks can then be outlined as:
\begin{enumerate}
 \item use a large-cone jet ($R\geq 0.8$) to cluster the whole top quark decay;
 \item apply an algorithm that looks inside the large-cone jet to try and recover the decay products of the top quark (top tagger);
 \item use jet substructure variables to discriminate between real top decays and other processes.
\end{enumerate}

\subsection{Jet grooming techniques}
The starting point of top quark taggers is the clustering of jets with a large radius ($\geq0.8$), however these jets tend to collect a lot of soft QCD radiation. Dedicated algorithms, called ``jet groomers'' are needed to resolve the hard decay products removing soft and wide angle radiation.

One of these algorithms is called ``jet trimming'' \cite{Krohn:2009th}, and consist in the following steps:
\begin{enumerate}
 \item inside the large-cone jet, cluster subjets of radius $R_{sub}$ using the $k_t$ algorithm
 \item reject the soft subjets that do not satisfy the condition $p_T$(subjet)$/p_T$(jet)$>f$.
\end{enumerate}
The optimal values of the parameters of the algorithm ($R_{sub}$ and $f$) are determined through Monte Carlo simulations.
Dedicated studies using full-detector simulation \cite{Aad:2013gja} show an improvement in performance of the reconstructed jet mass variable.

\section{Top quark tagging algorithms}
This section contains a summary description of the most widely used algorithms by the CMS \cite{cmsref} and ATLAS \cite{atlasref} experiments at the LHC.

\subsection{The CMSTopTagger algorithm}
The CMSTopTagger algorithm uses a Cambridge-Aachen jet with a radius of 0.8 as a starting point. A detailed description of the algorithm can be found in Ref.~\cite{Kaplan:2008ie}. The CMSTopTagger has been commissioned by the CMS Collaboration \cite{CMS:2014fya}.

\subsection{The HEPTopTagger algorithm}
The HEPTopTagger algorithm uses a Cambridge-Aachen jet with a radius of 1.5 as a starting point. Given the larger radius of the starting jet, this algorithm is able to cluster top decays with a smaller boost with respect to the other algorithms. A detailed description of the algorithm can be found in Ref.~\cite{Plehn:2010st}.
The HEPTopTagger has been commissioned by both the ATLAS and CMS Collaborations \cite{TheATLAScollaboration:2013qia}, \cite{CMS:2014fya}. 

The tagger's efficiency drops at very high $p_T$ because the initial clustering radius is too large and collects too many particles from the underlying event or pileup.
Thus a series of improvements have been developed leading to the Multi-R HEPTopTagger. Namely, the algorithm now tests different initial clustering radii and finds the optimal one. The optimal radius in function of jet $p_T$ can be used as a tagging variable.

\subsection{Calibration and data-driven corrections}
The usage of top tagging algorithms from large-cone jets poses a problem in the calibration of such tools for usage in real data.

The strategy adopted by the CMS Collaboration consists in deriving data--simulation correction factors computed using a ``tag and probe'' method in a semileptonic $t\bar{t}$ enriched selection. The idea is to ``tag'' an event by identifying a leptonic top, then probe for the tagger's efficiency on the hadronic side of the $t\bar t$ event.
The comparison of efficiencies derived in data and simulation gives the correction factors, measured in function of $\eta$, $p_T$, and Monte Carlo generators \cite{CMS:2014fya}.

The ATLAS Collaboration uses a different approach: for subjets of the HEPTopTagger a dedicated calibration is performed for different R subjets. For large-cone jets, the mass scale is validated by comparing the ratio of calo-jet mass divided by its correspondent track-jet mass in data and simulation \cite{Aad:2013gja}.

\subsection{Shower deconstruction tagger}
The starting point of the shower deconstruction tagger is the decomposition of the large-cone jet in small jets with radii between 0.1 and 0.3 (microjets). A detailed description of the algorithm can be found in Ref.~\cite{Soper:2012pb}.
Comparison of data with Monte Carlo simulations shows a reliable behavior of the discriminant and the kinematic variables related to the microjets \cite{shode}.

\section{Further jet substructure techniques}
\subsection{$B$-tagging in boosted topologies}
$B$-jet tagging is an essential tool in many physics analyses to distinguish the interesting signal containing top quarks from the multijet background.

$B$-tagging algorithms use information from jet tracks (especially the impact parameter of the track), and secondary vertices. The variables are then combined in a discriminator obtained from a neural network (MV1 tagger used by the ATLAS Collaboration) or a likelihood function (combined secondary vertex, CSV used by the CMS Collaboration).

In the boosted regime, the decay products of the hard scatter process have a small angular separation, in addition to this there is a higher probability of contamination with tracks from light flavor jets. These factors lead to a degraded performance of $b$-tagging algorithms.

The ATLAS Collaboration developed a series of improvement to the current $b$-tagging algorithm \cite{atlbtag1}, \cite{atlbtag2}. The first of these improvements regards the addition of new variables with more discrimination power between real $b$-jets and misidentified $b$-jets. The neural network used to derive the final discriminator variable is then retrained using samples enriched in high-$p_T$ $b$-jets. Considering a boosted topology, for a given $b$-tagging efficiency, the new algorithm yields a light flavor rejection rate that is, on average, twice that of the old algorithm.

The second point of improvement in the ATLAS $b$-tagging framework, is the usage of different input jets collections to the algorithm. Jets with a smaller cone are able to resolve better the boosted decay products, in addition to this, track jets provide a better jet direction resolution.

The CMS Collaboration uses two different approaches for $b$-tagging in boosted topologies \cite{CMS:2013vea}.
The first approach consist in running the CSV algorithm on the large-cone jets used to cluster the decay product of the hadronic top quark.

The second approach consist in running the CSV algorithm directly on the subjets found by a top quark tagging algorithm.
The second approach represent a more natural way to address the issue of $b$-tagging in top jets, as only one of the subjets is expected to come from the decay of a b quark. A performance assessment of the two approaches confirms that, for top jet identification, the subjet $b$-tagging approach is the most performing both in medium and very high boost scenarios.
Other improvements to the CSV algorithm have been implemented in the $b$-tagging framework \cite{newbtag}: among them there is the usage of the inclusive vertex finder (IVF) to search for secondary vertices.

\subsection{Other variables}
A number of other jet substructure variables provide discrimination power between jets coming from high-$pT$ top quark decays and other processes.

The N-subjettiness ($\tau_N$) \cite{Thaler:2010tr}, used by the CMS and ATLAS Collaborations, aims at describing how well a jet of radius $R_{jet}$ can be described as containing N or fewer $k_t$ subjets.
The variable is defined as:
\[{\tau_N=\frac{1}{d_0} \sum_k p_{T,k} {\rm min} \{ \Delta R_{1,k},\Delta R_{2,k},\cdots, \Delta R_{N,k} \}}\]
where $k$ runs over all the jet constituents and $d_0 = \sum_k p_{T,k}R_{jet}$.
The algorithm computes the $p_T$ weighted average of minimum $\Delta R$(jet constituent, subjet axis)$/R_{jet}$. For boosted hadronic top quarks the variable ratios $\tau_3/\tau_2$ and $\tau_2/\tau_1$ are relevant.

The $k_t$ splitting scale variable ($\sqrt{d_{i,j}}$), used by the ATLAS Collaboration, describes how likely it is for a jet to be composed by a two or three prongs decay. The algorithm is describe in Ref.~\cite{TheATLAScollaboration:2013qia}.

Both of these algorithms have been commissioned on data collected by the LHC at a center of mass energy of 8 TeV \cite{CMS:2014fya}, \cite{TheATLAScollaboration:2013qia}. The variables have been used in a number of physics analyses.

\subsection{The semi-resolved case and W-tagging}
In medium boost regimes ($p_T$(jet)$\approx m_{\text{top}}$) angular separation between the decay products of a top quark might be big enough that a large-cone jet is not able to cluster the whole decay. In these cases it is possible to cluster the $b$-jet and the hadronic decay products of the W boson in two separate jets. For the $b$-jet reconstruction, standard size jets ($R=0.4-0.5$) and standard $b$-tagging techniques can be used. The hadronic W boson decay can then be clustered separately in a large-cone jet (W-jet).

To distinguish jets originated from boosted W boson decay from other processes a class of algorithms called W-taggers is used.
W-jets are a benchmark topology for jet substructure studies, so a large amount of techniques were developed in the past few years \cite{CMS:2014fya}, \cite{Khachatryan:2014vla}, \cite{CMS:2014joa}, \cite{atlw}. One of the most used techniques used to identify large cone jets coming from W bosons consists in the following steps:
\begin{enumerate}
 \item use a jet grooming algorithm;
 \item compute the jet mass from the groomed jet and apply a cut on the groomed jet mass;
 \item eventually use a jet substructure variable to select two-prong decays (e.g. N-subjettiness).
\end{enumerate}

\subsection{Pileup mitigation techniques for Run 2}
Pileup represents a major issue in the reconstruction and energy measurement of large-cone jets,  commonly used for top quark tagging. Jet substructure variables built starting from the jet components are also affected by the same problem. Many standard techniques to treat pileup contamination exist already.
However, dedicated techniques are being developed to address pileup in the forthcoming data taking period at the LHC.
One of these techniques is pileup per particle identification (PUPPI) \cite{Bertolini:2014bba}.

This new approach consist in assigning a weight to each reconstructed object in the detector (by scaling its momentum) based on pileup event properties and tracking information. The novelty of the approach consist in acting directly on the input component of the jet clustering algorithm; the immediate consequence of this fact is that all the jet substructure variables computed using the jet components as input are automatically corrected by the procedure \cite{CMS:2014ata}.

\section{Algorithm comparison}
\begin{figure}
\centering
\includegraphics[height=0.22\textheight]{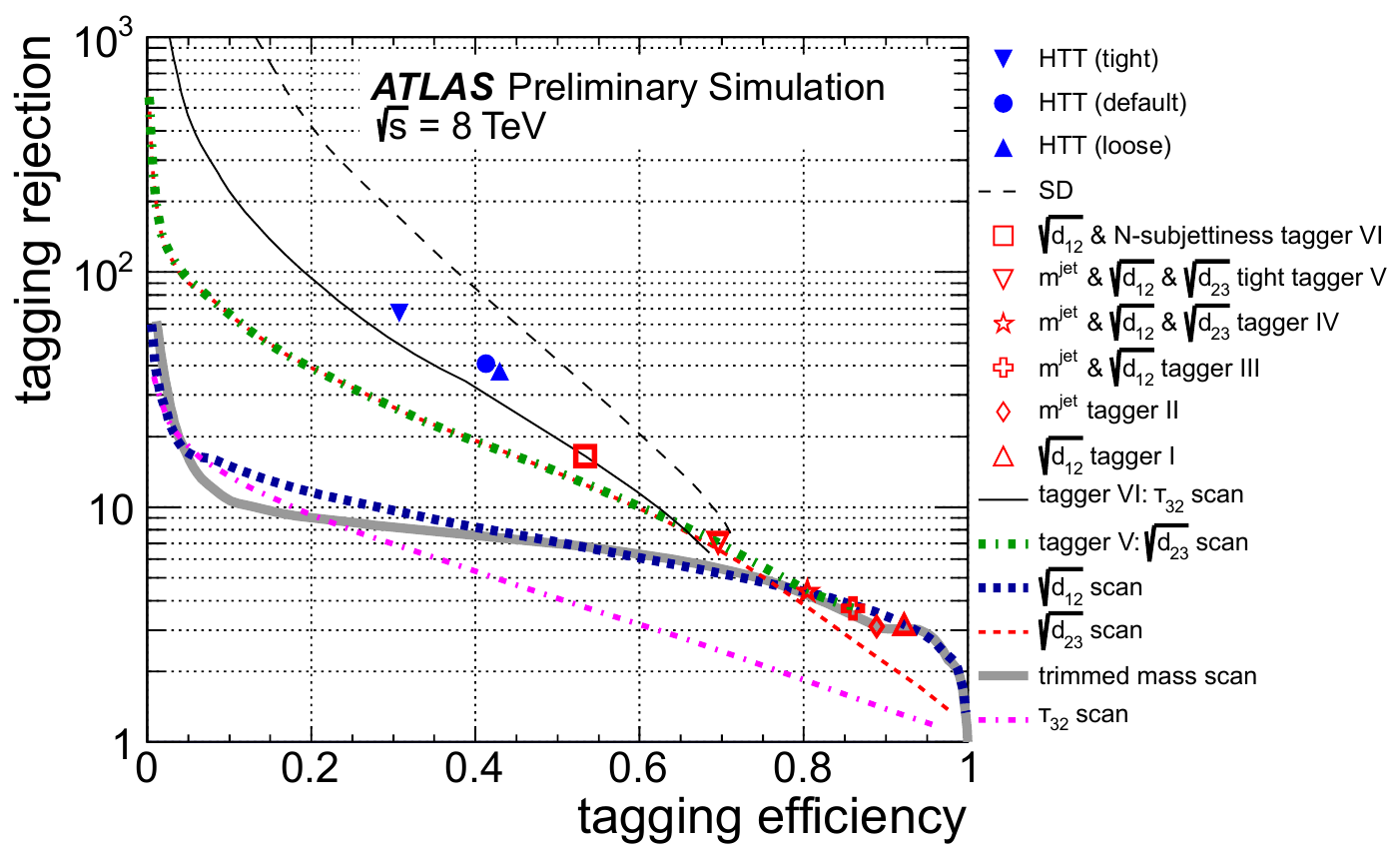}\\
\includegraphics[height=0.209\textheight]{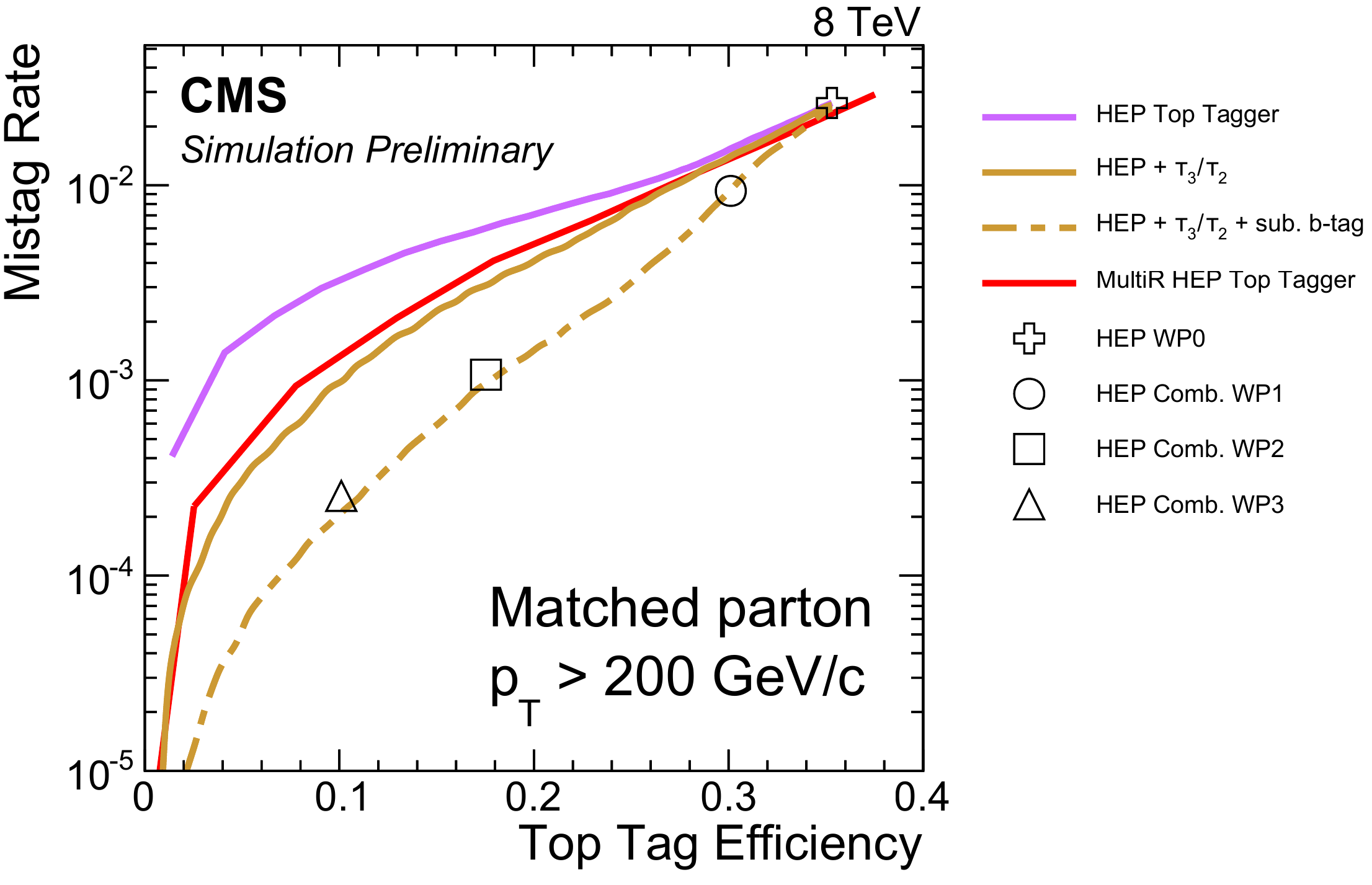}
\includegraphics[height=0.209\textheight]{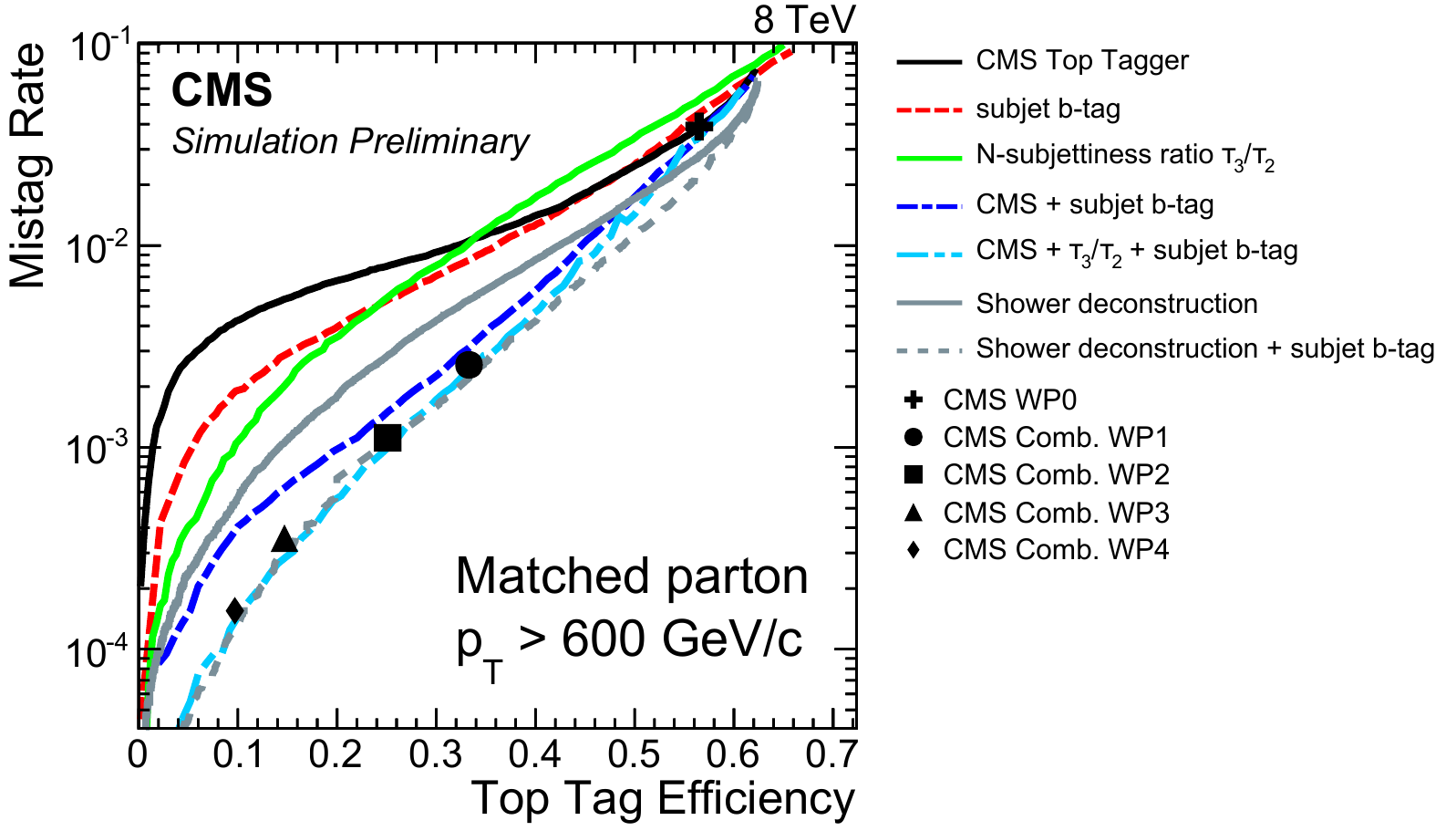}
\caption{Top row: top tagging performance in ATLAS using trimmed anti-$k_t$ R=1.0 jets and Cambridge-Aachen R=1.2 jets (for the HEPTopTagger), $p_T$(jet)$>$ 550 GeV \cite{shode}. Bottom row: top tagging performance in CMS using R=1.5, $p_T$(matched parton)$>$ 200 GeV jets (left) and R=0.8, $p_T$(matched parton)$>$ 600 GeV jets (right) \cite{CMS:2014fya}.}\label{fig:perf}
\end{figure}

The various top tagging algorithms and jet substructure variables can be compared by computing performance curves (also called ROC curves sometimes), plotting the top misidentification probability in function of the tagger's efficiency. A continuous line can be obtained for each tagger by varying the parameters of the algorithm. Different tagging algorithms and jet substructure variables can be combined together to obtain a better discrimination power. Performance comparison is shown for ATLAS and CMS in Fig.~\ref{fig:perf}.




\section{Conclusions}
Boosted top quark tagging and, more in general, jet substructure techniques are a very active field of research with many new theoretical and experimental development presented every year. These tools are now widely used in searches for physics beyond the standard model with top quarks in the final state. Such techniques will be even more relevant during Run 2 of the LHC.

Studies are ongoing to assess the performance of the taggers under the pileup scenarios that will be found during the next run of the LHC and to better understand the systematic uncertainties associated to the use of top quark tagging in physics analyses.

\end{document}